\documentclass[]{aastex63}
\usepackage{multirow}
\usepackage{graphicx}
\usepackage{threeparttable}
\usepackage{xcolor}
\usepackage{harveyballs}

\makeatletter
\global\let\tikz@ensure@dollar@catcode=\relax
\makeatother

\newcommand\Lucy{{\sl Lucy}}

\newcommand\dg{$^{\circ}$}

\newcommand\x{$\times$}
\newcommand\um{$\mu$m}
\newcommand\kgm{kg/m$^{3}$}

\received{December 18, 2021}
\accepted{February 24, 2021}
\submitjournal{PSJ}

\shorttitle{\Lucy\ Mission Concept of Operations and Instruments}
\graphicspath{{./}{figures/}}

\begin{document}

\title{\Lucy\ Mission to the Trojan Asteroids: Instrumentation and \color{black}Encounter \color{black} Concept of Operations}

\correspondingauthor{Catherine Olkin}
\email{colkin@boulder.swri.edu}

\author[0000-0002-5846-716X]{Catherine B. Olkin}
\affiliation{Southwest Research Institute \\
1050 Walnut Street, Suite 300 \\
Boulder, CO 80302, USA}

\author[0000-0001-5847-8099]{Harold F. Levison}
\affiliation{Southwest Research Institute \\
	1050 Walnut Street, Suite 300 \\
	Boulder, CO 80302, USA}

\author{Michael Vincent}
\affiliation{Southwest Research Institute \\
	1050 Walnut Street, Suite 300 \\
	Boulder, CO 80302, USA}

\author[0000-0002-6013-9384]{Keith S. Noll}
\affiliation{NASA Goddard Spaceflight Center\\
8800 Greenbelt Rd. \\
Greenbelt, MD 20771}

\author{John Andrews}
\affiliation{Southwest Research Institute \\
	1050 Walnut Street, Suite 300 \\
	Boulder, CO 80302, USA}

\author{Sheila Gray}
\affiliation{Lockheed Martin Corp.\\
12257 South Wadsworth Blvd.\\
Littleton, CO 80125}

\author{Phil Good}
\affiliation{Lockheed Martin Corp.\\
12257 South Wadsworth Blvd.\\
Littleton, CO 80125}

\author[0000-0003-2548-3291]{Simone Marchi}
\affiliation{Southwest Research Institute \\
	1050 Walnut Street, Suite 300 \\
	Boulder, CO 80302, USA}

\author[0000-0001-9625-4723]{Phil Christensen}
\affiliation{Arizona State University\\
Tempe, AZ 85281}

\author[0000-0002-6829-5680]{Dennis Reuter}
\affiliation{NASA Goddard Spaceflight Center\\
8800 Greenbelt Rd. \\
Greenbelt, MD 20771}

\author[0000-0003-0951-7762]{Harold Weaver}
\affiliation{Johns Hopkins University Applied Physics Laboratory\\
7651 Montpelier Rd.\\
Laurel, MD 20723}

\author[0000-0003-3479-856X]{Martin P\"{a}tzold}
\affiliation{RIU-PF at Cologne University\\
Cologne, Germany}

\author[0000-0002-2006-4074]{James F. Bell III}
\affiliation{Arizona State University\\
Tempe, AZ 85281}

\author[0000-0001-8675-2083]{Victoria E. Hamilton}
\affiliation{Southwest Research Institute \\
	1050 Walnut Street, Suite 300 \\
	Boulder, CO 80302, USA}

\author[0000-0002-8379-7304]{Neil Dello Russo}
\affiliation{Johns Hopkins University Applied Physics Laboratory\\
7651 Montpelier Rd.\\
Laurel, MD 20723}

\author[0000-0003-4641-6186]{Amy Simon}
\affiliation{NASA Goddard Spaceflight Center\\
8800 Greenbelt Rd. \\
Greenbelt, MD 20771}

\author{Matt Beasley}
\affiliation{Southwest Research Institute \\
	1050 Walnut Street, Suite 300 \\
	Boulder, CO 80302, USA}

\author[0000-0002-8296-6540]{Will Grundy}
\affiliation{Lowell Observatory\\
1400 Mars Hill Rd.\\
Flagstaff, AZ 86001}

\author[0000-0003-1869-4947]{Carly Howett}
\affiliation{Southwest Research Institute \\
	1050 Walnut Street, Suite 300 \\
	Boulder, CO 80302, USA}

\author[0000-0003-4452-8109]{John Spencer}
\affiliation{Southwest Research Institute \\
    1050 Walnut Street, Suite 300 \\
	Boulder, CO 80302, USA}
 
\author[0000-0003-4810-7352]{Michael Ravine}
\affiliation{Malin Space Science Systems \\
    5880 Pacific Center Blvd. \\
    San Diego, CA 92121}

\author{Michael Caplinger}
\affiliation{Malin Space Science Systems \\
    5880 Pacific Center Blvd. \\
    San Diego, CA 92121}



\begin{abstract}

The \Lucy\ Mission accomplishes its science during a series of five flyby encounters with seven Trojan asteroid targets. This mission architecture drives a concept of operations design that maximizes science return, provides redundancy in observations where possible, features autonomous fault protection and utilizes onboard target tracking near closest approach. These design considerations reduce risk during the relatively short time-critical periods when science data is collected. The payload suite consists of a color camera and infrared imaging spectrometer, a high-resolution panchromatic imager, and a thermal infrared spectrometer. The mission design allows for concurrent observations of all instruments. Additionally, two spacecraft subsystems will also contribute to the science investigations: the Terminal Tracking Cameras will obtain wide field-of-view imaging near closest approach to determine the shape of each of the Trojan targets and the telecommunication subsystem will carry out Doppler tracking of the spacecraft to determine the mass of each of the Trojan targets.   

\end{abstract}



\section{Introduction} \label{sec:intro}

\begin{figure}[hb]
	\centering
	\includegraphics[scale=0.28]{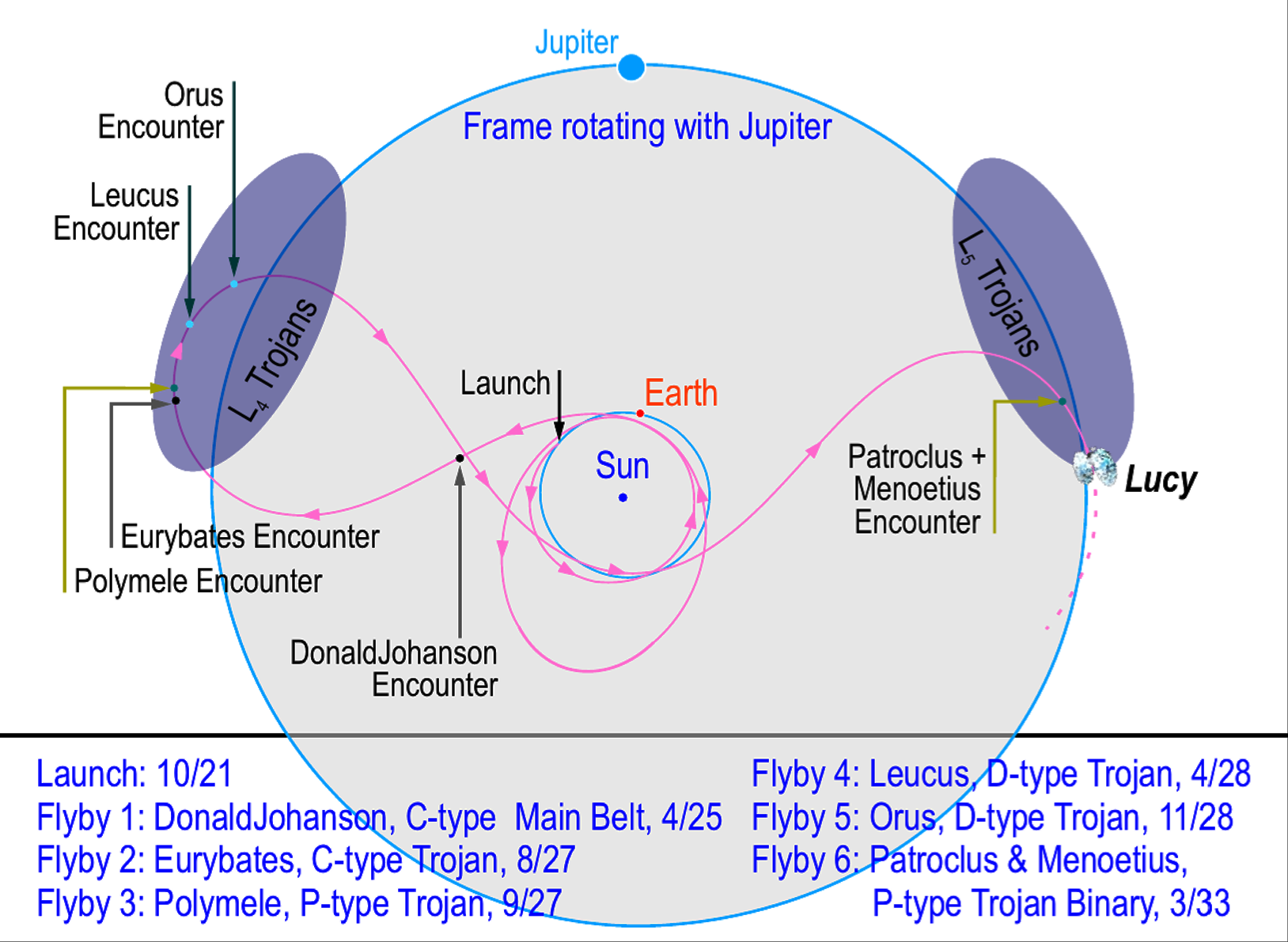}
	\color{black}
	\caption{The \Lucy\ trajectory is shown in a frame rotating with Jupiter. The launch window opens on October 16, 2021. After two Earth gravity assists (in 2022 and 2025) that raise the aphelion, \Lucy\ transfers to the $L_4$ Trojan cloud where the spacecraft encounters Eurybates and Queta, Polymele, Leucus, and Orus between August 2027 and November 2028. After a return to Earth for another gravity assist in December 2030, \Lucy\ heads for the $L_5$ cloud where it encounters the Patroclus-Menoetius binary (PMB) in March 2033.}
	\color{black}
	\label{fig:pretzel}
\end{figure}

The \Lucy\ mission consists of five flybys of Trojan asteroids to investigate the differences in surface and internal properties across the population of Trojan asteroids. From these five encounters we will be able to observe seven Trojan asteroids: (3548) Eurybates and its small satellite Queta, (15094) Polymele, (11351) Leucus, (21900) Orus, (617) Patroclus and (617) Meneotius (Fig. \ref{fig:pretzel}). Two of the flybys will encounter multiple Trojan asteroids. The first \Lucy\ Trojan flyby will be of Eurybates and its recently discovered small satellite \citep{Noll+2020b} and the last encounter is of the near-equal size binary system: Patroclus and Meneotius. \Lucy\ will also fly by a Main Belt asteroid target of opportunity, (52246) Donaldjohanson, prior to reaching the Trojans, and will use this encounter to test operations.

During the flybys, the spacecraft is moving relative to the Trojan asteroids with a velocity of $\sim$6-9 km/s making time a critical resource. The mission is designed to maximize the data collected around closest approach which requires efficiency in observing the Trojan asteroids. This paper describes key considerations for the concept of operations and instrumentation of the \Lucy\ mission. More background, including the science objectives of the mission, can be found in the companion paper by \cite{Levison+2021}.

\color{black}
\section{Trajectory} \label{sec:trajectory}

The early design of the Lucy mission focused on the development of the trajectory and target selection. The purpose of Lucy is to perform the first reconnaissance of the Trojan asteroids and to investigate the diversity of the population. The initial desire was to find a matched pair of objects that differed in spectral type and collisional history. The pair of Eurybates and (21900) 1999 VQ$_{10}$, which we later named Orus, nicely met this requirement. The objects have similar diameters and orbital inclinations and eccentricities. Orus is a typical, spectrally red D-type Trojan. Eurybates, on the other hand, is a spectrally-neutral C-type that is rare in the Trojans (but very common in the Outer Main Asteroid Belt). Moreover, it is the largest member of a disruptive collisional family, also a rarity in the Trojans. The combination of an object with two unusual characteristics and a “control” object provides a nearly ideal experimental setup to address the possible causal linkage between color and collisional processing that would have much broader applicability to the Trojan, Centaur, and Transneptunian populations. 

After finding the Eurybates and Orus pair, both in the $L_4$ cloud, an extension of the trajectory to the $L_5$ cloud after another Earth gravity assist showed that it was possible to intercept the high-inclination (617) Patroclus-Menoetius binary when it was fortuitously close to the ecliptic plane. The binary is spectrally intermediate between Eurybates and Orus and the survival of the binary requires moderate limits on collisions in this system. The combination of Eurybates, Orus, and Patroclus-Menoetius gives a rich sampling of the diversity of the Trojans and was the baseline mission proposed to NASA.

During phase A, further refinements to flight dynamics made it feasible to add two more $L_4$ targets to the Lucy baseline. Leucus is a second D-type object which is notable for its very slow rotation and large amplitude lightcurve. Polymele is a P-type object and the smallest Trojan that Lucy will encounter other than the small satellite of Eurybates.

Although not a part of the baseline or threshold science missions, Lucy will also fly by a small Main Belt Asteroid on its way to the $L_4$ cloud. (52242), which we named Donaldjohanson for the discoverer of the Lucy fossil that inspired the name for the \Lucy\ mission, is a C-type object and a small member of the 280 Myr-old Erigone collisional family \citep{Vokrouhlicky+2006}. The mission will use this flyby to test out sequences that will be used for the Trojan encounters, but scientific data will be acquired to the extent possible. 

\color{black}

\section{Science Payload} \label{sec:payload}

The Science Payload on \Lucy\ is comprised of three instruments: L'LORRI (\Lucy\ LOng Range Reconnaissance Imager; a high resolution panchromatic camera), L'Ralph (a color imager and short-wavelength infrared imaging spectrometer) and L'TES (\Lucy\ Thermal Emission Spectrometer; a thermal infrared spectrometer). The \textit{L'} indicates that these are the \Lucy\ versions of these instruments rather than their predecessor instruments.

L'Ralph has two different focal planes: MVIC (Multi-spectral Visible Imaging Camera) and LEISA (Linear Etalon Imaging Spectral Array) and in many ways these can be considered operationally as two independent instruments despite the fact that they share part of their optical train. In addition to the three scientific instruments, two spacecraft subsystems will also provide data to satisfy our science objectives as detailed below in Section \ref{sec:systems}. 

The instrument and spacecraft subsystem specifications are given in Table \ref{tab:InstrumentSpecs}. The spacecraft and instruments are still in development so the values in Table \ref{tab:InstrumentSpecs} reflect the requirements. For the mass and power, these values are the NTE (Not To Exceed) numbers allocated for that instrument or subsystem. MVIC and LEISA are both part of L'Ralph and it is not possible to break out a mass for MVIC and a mass with LEISA (the same is true for power). The Ralph instrument mass (35.5 kg) and power (30.0 W) are shown instead. 

\begin{deluxetable}{lcccccc}
	\tabletypesize{\scriptsize}
	\label{tab:InstrumentSpecs}	
	\tablecolumns{8}
	\tablewidth{0pt}
	\tablecaption{\Lucy\ Instrument Specifications}
	\tablehead{	\colhead{} & \colhead{L'LORRI}& \colhead{L'TES}& \colhead{L'Ralph/MVIC} & \colhead{L'Ralph/LEISA} & \colhead{TTCam}}
	\startdata
	Mass, kg & 14.0 & 7.7 & \multicolumn{2}{c}{37.0}  & 2.96 \\
	Power, W & 12.4 & 17.6 & \multicolumn{2}{c}{30.0}  & 10 \\
	FOV,$^{\circ}$  & 0.29 & 0.57 & 8.3 & 1.4x3.4 & 11x8.2 \\
	iFOV, $\mu rad$/pixel  & 5.0 & N/A & 29 & 80 & 74 \\
	Spectral Range, $\mu m$ & 0.45-0.85 & 6-75 & 0.38-0.92 & 1.0-3.8 & 0.425-0.675 \\
	Spectral Resolution & 400 nm & 10 cm$^{-1}$ & 47-550 nm & 10 nm & 250 nm \\
	\enddata
\end{deluxetable}


\color{black}A mapping of the science requirements to the instruments and spacecraft subsystems is shown in Table \ref{tab:Mapping}. 

\begin{longrotatetable}
\begin{table*}[]
\color{black}
  \centering
  \caption{\bf Mapping Lucy's Science Requirements to Instruments}
  \begin{tabular}{ll|c|ccccccc}
\hline
No & Descriptor & Baseline Requirement & LOR & MVI & T2C &LEI &TES & RS\\
\hline
\hline
    1 & Targets & Patroclus,Meneotius, Eurybates, Leucus, Polymele, and Orus  & \harveyBallNone & \harveyBallNone& \harveyBallNone& \harveyBallNone& \harveyBallNone& \harveyBallNone \\
    2 & Shape and Geology &  Pan images: full rotation spaced by 1/25 to 1/13 of a rotation  & \harveyBallHalf & \harveyBallHalf& \harveyBallFull & \harveyBallNone & \harveyBallNone& \harveyBallNone \\
    3 & Shape and Geology & Pan images: series of phase angles separated by 15\dg - 25\dg  & \harveyBallHalf & \harveyBallHalf& \harveyBallFull & \harveyBallNone& \harveyBallNone& \harveyBallNone \\
    4	& Elevation Models &  Pan images: area $\geq$ 100 km$^2$; resolution $\leq$ 200 m, two stereo emission angles   & \harveyBallFull & \harveyBallHalf& \harveyBallFull & \harveyBallNone& \harveyBallNone& \harveyBallNone \\
    5 & Landform Degr. & Pan images: area $\geq$ 500 km$^2$; equator to 60\dg lat.; resolution $\leq$ 100 m & \harveyBallFull & \harveyBallFull & \harveyBallFull & \harveyBallNone& \harveyBallNone& \harveyBallNone \\
    6 & Craters & Pan images: area $\geq$ 700 km$^2$; resolve craters d $>$ 7 km & \harveyBallHalf & \harveyBallFull & \harveyBallFull & \harveyBallNone& \harveyBallNone& \harveyBallNone \\
    7 & Craters  & Pan images: area $\geq$ 10 km$^2$; resolve craters d $>$ 70 m & \harveyBallFull &  \harveyBallQuarter &  \harveyBallQuarter & \harveyBallNone& \harveyBallNone& \harveyBallNone \\
    8 & Satellite Search & Search: satellites d $\geq$ 2 km, $p_v > 0.04$ within R$_{Szebehely}$  & \harveyBallFull & \harveyBallHalf& \harveyBallNone & \harveyBallNone& \harveyBallNone& \harveyBallNone \\
    9 & Global Color & Color images: full rotation spaced by 1/6 to 1/3 of a rotation & \harveyBallNone & \harveyBallFull & \harveyBallNone& \harveyBallNone& \harveyBallNone& \harveyBallNone \\
    10 & Low-Res Color & Color images: area $\geq$ 700 km$^2$; resolution $\leq$ 1.5 km  & \harveyBallNone & \harveyBallFull & \harveyBallNone & \harveyBallNone& \harveyBallNone& \harveyBallNone \\
    11 & High-Res Color & Color images: area $\geq$ 150 km$^2$; resolution $\leq$ 600 m   & \harveyBallNone & \harveyBallFull & \harveyBallNone& \harveyBallNone& \harveyBallNone& \harveyBallNone \\
    12 & NIR Range & Spec: spectral range 1.0-3.8 \um & \harveyBallNone & \harveyBallNone& \harveyBallNone & \harveyBallFull& \harveyBallNone& \harveyBallNone \\
    13 & NIR Perf. & Spec: detect features with spectral depth $\geq$ 4\% and width of $\geq$ 70 nm  & \harveyBallNone & \harveyBallNone& \harveyBallNone & \harveyBallFull& \harveyBallNone& \harveyBallNone \\
    14 & Compositional Variation & Spec: full rotation spaced by 1/6 to 1/3 of a rotation & \harveyBallNone & \harveyBallNone& \harveyBallNone & \harveyBallFull& \harveyBallNone& \harveyBallNone \\
    15 & Composition & Spec: resolution (r) and areal coverage (A) satisfy $r \leq 2(A/1470.6)^{0.473}$ km & \harveyBallNone & \harveyBallNone& \harveyBallNone & \harveyBallFull& \harveyBallNone& \harveyBallNone \\
    16 & Mass &  Targets: mass accuracy $\leq$ 25\%, for $\rho\geq 1000$ \kgm & \harveyBallNone & \harveyBallNone& \harveyBallNone & \harveyBallNone& \harveyBallNone& \harveyBallFull \\
    17 & Thermal & Thermal: 1 unilluminated surface; 3 sunlit with 1 at $<$ 30\dg from subsolar point & \harveyBallNone & \harveyBallNone& \harveyBallNone & \harveyBallNone& \harveyBallFull& \harveyBallNone \\
    \hline
  \end{tabular}
  \label{tab:Mapping}
  \tablecomments{The instruments abbreviations are L'LORRI (LOR) a panchromatic high-resolution imager, MVIC (MVI) a panchromatic and color imager that is part of L'Ralph, the Terminal Tracking Cameras (T2C), LEISA (LEI) the infrared imaging spectrograph that is part of L'Ralph, L'TES (TES) a thermal emission spectrograph, and Radio Science (RS) which uses the telecommunication subsystem to measure the doppler shift over a flyby. The circles are fully filled in for instruments that can completely accomplish the requirement. The half-filled circle indicates the instrument partially contributes to the requirement. The quarter-filled circle indicates that this instrument provides a degraded backup for the requirement. Finally, an open circle indicates that the instrument does not contribute to the requirement.}
\end{table*}
\end{longrotatetable}
\color{black}

\subsection{L'LORRI}

The L'LORRI instrument provides the highest-resolution imaging for the \Lucy\ mission. The panchromatic images will address 7 of our 17 Level 1 science requirements \color{black}(numbers 2-8 in Table \ref{tab:Mapping}) \color{black} including searching for satellites, determining the size frequency distribution of craters, stereo imaging and shape reconstruction. The L'LORRI instrument will also be used for optical navigation to our targets.

\begin{figure}[htbp]
	\centering
	\includegraphics[keepaspectratio,width=5 in]{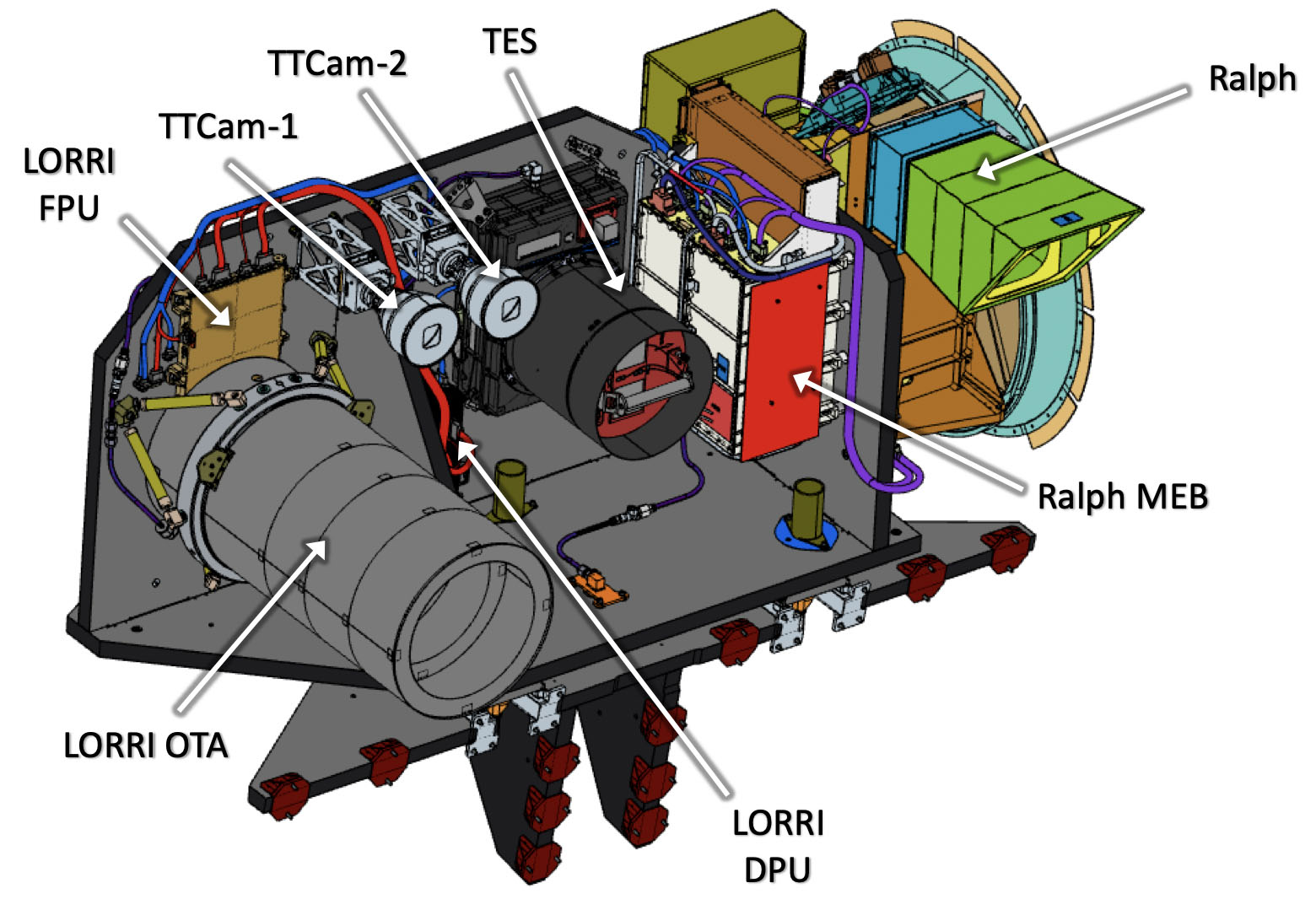}
	\caption{The Instrument Pointing Platform (IPP). From left to right, the instruments are L'LORRI with its associated FPU (Focal Plane Unit) and Digital Processing Unit (DPU), two Terminal Tracking Cameras (TTCam-1 and TTCam-2), L'TES, and L'Ralph with its associated MEB (main electronics box).}
	\label{fig:IPP}
\end{figure}

The instrument is being built at the Johns Hopkins University Applied Physics Lab and Dr.~Harold Weaver is the Instrument Principal Investigator. The instrument design is derived from the LORRI instrument \citep{Cheng+2008} on New Horizons. L'LORRI uses the same detector and has the same optical design as New Horizons LORRI. The primary mirror has a diameter of 20.8 cm and the system has a focal length of 262 cm and the detector is a 1024\x1024 thinned back-illuminated frame transfer CCD from Teledyne e2v. Each pixel subtends 5 $\mu$rad and will have a point spread function with a full width half maxiumum (FWHM) of less than 15 $\mu$rad. Differences from the heritage instrument worth noting are the addition of redundant electronics, memory to store LORRI data, and the difference in the instrument accommodation. On New Horizons the LORRI instrument is inside of the spacecraft, but on \Lucy\ L'LORRI is mounted on an Instrument Pointing Platform (IPP) as shown in Figure \ref{fig:IPP}.

\begin{figure}[ht!]
	\centering
	\includegraphics[keepaspectratio,width=4 in]{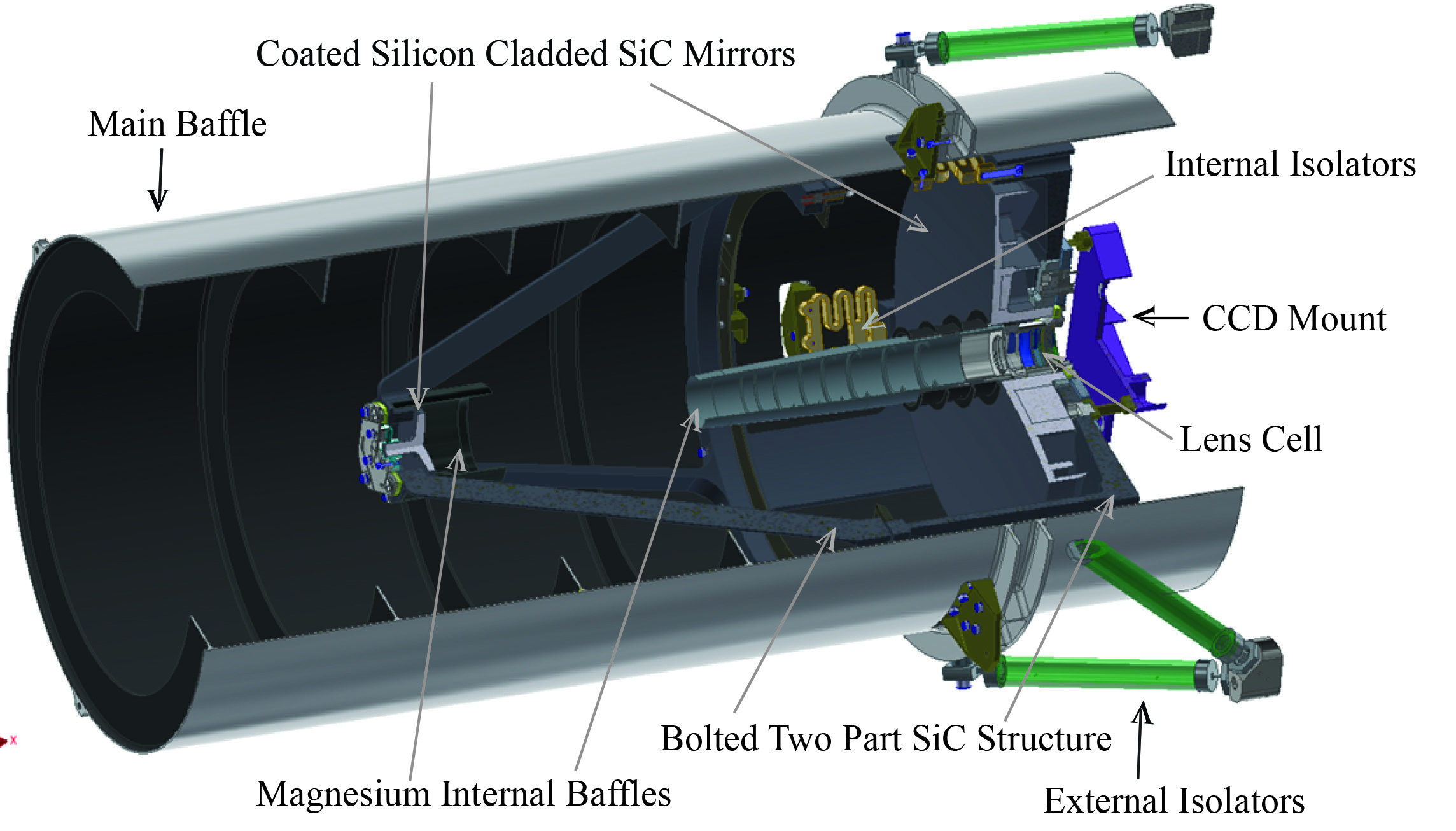}
	\caption{CAD drawing of the L'LORRI instrument. The LORRI telescope is a Ritchey-Chr\'{e}tien optical design with a SiC structure. The external isolators are used to mount the telescope to the IPP. \label{fig:LORRI_CAD}}
\end{figure}

The instrument has two modes of operation: 1\x1 mode (unbinned data) and 4\x4 mode (binning the pixels 4\x4 before being read out). The primary purpose for the 4\x 4 mode is to increase sensitivity for the satellite search. To accomplish the satellite search, L'LORRI is required to have a SNR $\geq$ 7 in 4\x4 mode for a 10 second exposure time on a V=15.8 magnitude source. For the high resolution imaging of a resolved source the instrument needs to have a SNR $\geq$ 25 in 100 ms exposure time for $I/F = 0.0014$ at 5.7 AU. The $I/F$ corresponds to a Trojan asteroid with a geometric albedo of 3\% at a phase angle of 82$^{\circ}$ with a phase curve of 0.04 mag/degree. The Lucy mission can accomplish the science objectives for each encounter on just one asymptote, preferably the one with the smaller solar phase angle as it has the most illuminated terrain. Across all encounters as seen in Table \ref{tab:EncounterCircumstances}, the minimum phase angles span from the smallest solar phase angle at Orus (54$^{\circ}$) to the largest value 82$^{\circ}$ for Polymele. Therefore, the Lucy mission set the 82$^{\circ}$ phase angle for the SNR requirement. 

L'LORRI also has a requirement to not saturate when observing a target with $I/F = 0.40$ that covers half the field of view. This requirement was driven from the experience of Dawn at Ceres which observed a  highly reflective surface in Occator crater \citep{DeSanctis+2016}.

\subsection{L'Ralph}

The L'Ralph instrument is basically two instruments in one: (i) a Multispectral Visible Imaging Camera (MVIC) and (ii) a Linear Etalon Imaging Spectral Array (LEISA). MVIC provides both panchromatic and color imaging of the Trojan asteroids to address or contribute to 14 of \Lucy's Level 1 science requirements and MVIC is solely responsible for the 3 color requirements. LEISA provides spectral information to map the surface composition of the Trojan asteroids addressing 4 of the Level 1 science requirements. \color{black}See Table \ref{tab:Mapping}. \color{black}

The instrument is being built at Goddard Spaceflight Center and Dr.~Dennis Reuter is the Instrument Principal Investigator. The instrument design is based on the Ralph instrument \citep{Reuter+2008} and the OVIRS instrument \citep{Reuter+2018}. 

The L'Ralph instrument has a 3-mirror anastigmat design f/6 with a 75 mm aperture. The telescope structure is composed from one Aluminum block to provide an athermal imaging system. \color{black}A beamsplitter \color{black} transmits the longer wavelength light to LEISA and reflects light short of $\sim$960 nm to MVIC. The instrument is passively cooled with a 20-inch diameter radiator that cools the LEISA detector to $\sim$100K. A new component of the L'Ralph instrument compared to its predecessors is a scan mirror assembly. The scan mirror is used to sweep the target across the Ralph focal planes to build up either visible images or infrared spectra.

\begin{figure}[htbp]
	\centering
	\includegraphics[keepaspectratio,width=4 in]{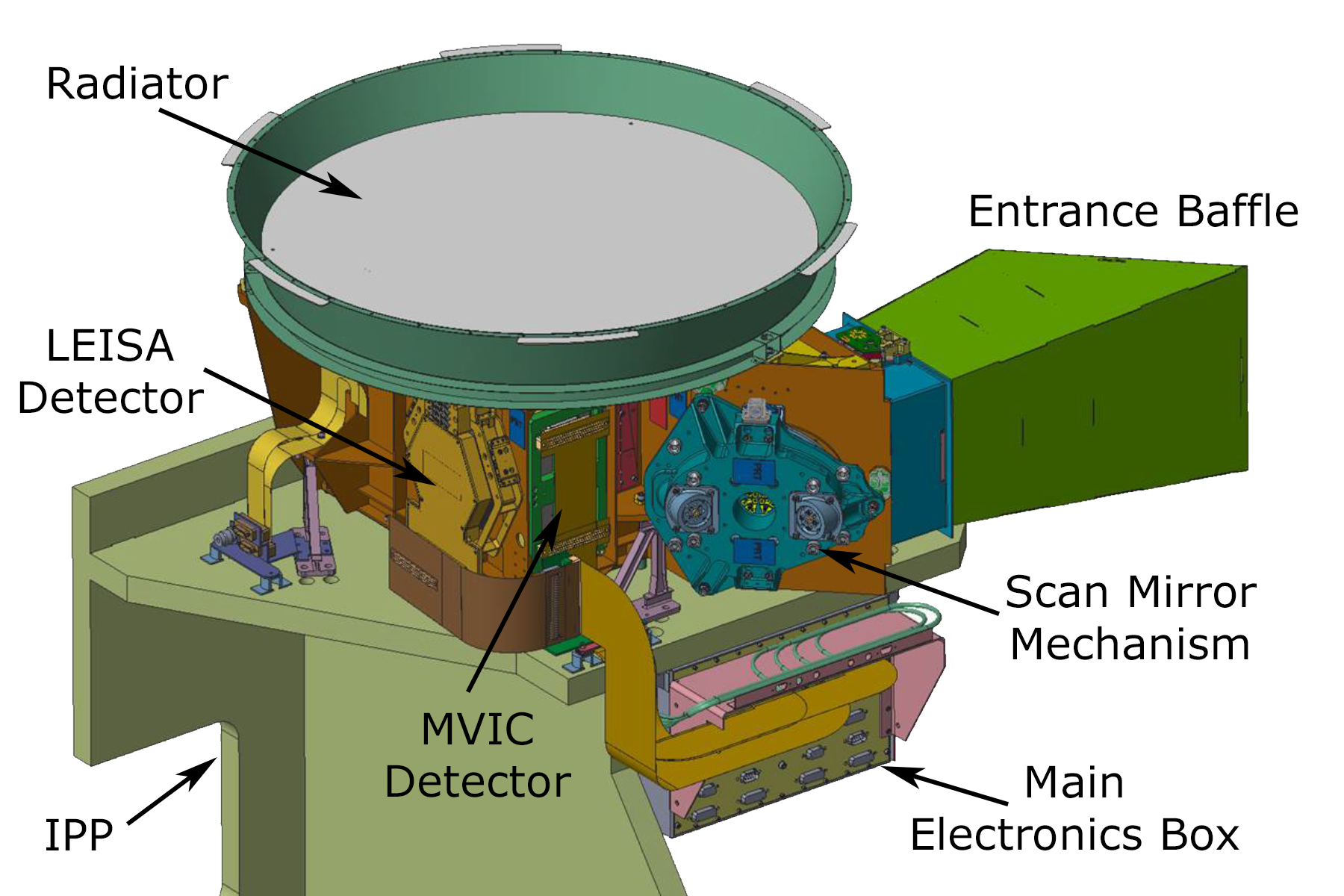}
	\caption{CAD drawing of the L'Ralph instrument mounted on the Instrument Pointing Platform (IPP). }
	\label{fig:RalphCAD}
\end{figure}

The MVIC focal plane consists of 6 Time Delay Integration (TDI) charge-coupled devices (CCD) on a single substrate supplied by STA (Semiconductor Technology Associates). There are 5 color channels and 1 panchromatic channel with passbands as shown in Table \ref{tab:filters}. With a TDI CCD, signal is built up as the scene is scanned over the focal plane in coordination with the charge being transferred between rows of the detector. The MVIC detectors are 5000\x64 pixels and each pixel subtends 29 $\mu$rad and have more than 50$\%$ of the encircled energy within 25 $\mu$rad. The FOV in cross track direction (5000 pixel side of the detector) is 8.3$^{\circ}$ and the \color{black}system \color{black} can be commanded to sum over 4, 8, 16, 32 or 64 TDI rows. The system is designed to achieve a SNR of at least 25 for our color observations and not to saturate when observing a scene with $I/F = 0.4$.

\begin{deluxetable}{lc}
	\tabletypesize{\scriptsize}
	\label{tab:filters}
	\tablewidth{0pt}
	\tablecaption{Filter Passbands}
	\tablehead{\colhead{Filter}&\colhead{Wavelength Range, nm }}
	\startdata
	Violet&377-489\\
	Green&478-525\\
	Orange&522-632\\
	Phyllosilicate&627-758\\
	Near-IR&753-914\\
	Pan&377-914\\
	\enddata
\end{deluxetable}

The LEISA detector is an H2RG \citep{Blank+2011} detector with 3 linear variable filters to disperse the light from 0.95 to 3.95 \um. \color{black} The H2RG is an infrared focal plane array with HgCdTe detector material. This detector is made by Teledyne Technologies Company. \color{black} For each image one direction is spatial and the other is spectral. The scan mirror sweeps the image of the target across the LEISA detector in the spectral direction to build up a image cube which has a full spectrum of each observed surface element of the Trojan. In the spatial direction, the pixels subtend 80 $\mu$rad and will have a more than 50$\%$ of the encircled energy within 68 $\mu$rad.
  
The spectral resolution of LEISA is 10 nm. This resolution allows for \Lucy\ to sample the 70 nm wide polycyclic aromatic hydrocarbons (PAH) absorption feature at 3.29 \um\  which is a class of organic molecules of interest and accessible in the LEISA spectral range. LEISA will achieve a SNR greater than 25 from 1 to 3.4 \um. For the rest of the required spectral range (from 3.4 - 3.8 \um), there is not a required minimum SNR.

The primary science modes of L'Ralph are 5-color imaging with MVIC, color and panchromatic imaging with MVIC, and observing with LEISA. The \Lucy\ spacecraft can take simultaneous observations with all our instruments but L'Ralph can not simultaneously observe with both MVIC and LEISA.  

\subsection{L'TES}

The L’TES instrument measures the thermal infrared emission from each Trojan asteroid to obtain the temperature of the asteroid’s surface. These observations address one of the Level 1 science requirements: determining the thermal inertia of the surface. The L’TES is a near-copy of the OTES instrument on OSIRIS-REx \citep{Christensen+2018}, where OTES is used to derive the surface composition and thermal inertia of the asteroid Bennu. However, because the Trojan asteroids at 5 AU are much colder than Bennu, the \Lucy\ mission does not plan to use L’TES to derive surface composition. Instead, L’TES will be used primarily to infer regolith properties.

\begin{figure}[ht]
\begin{center}
	\includegraphics[scale=0.4]{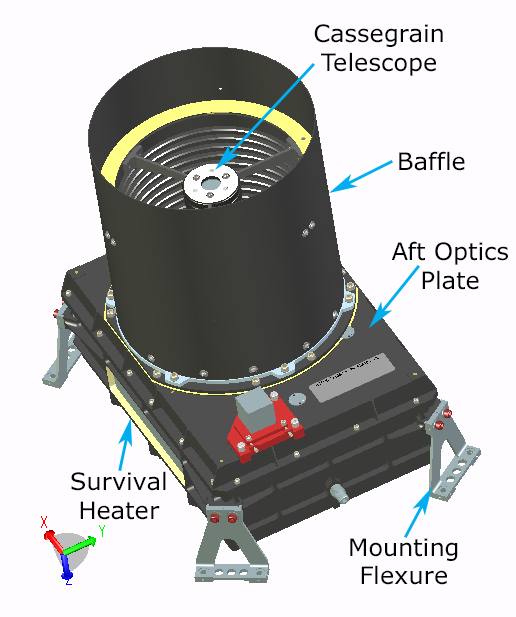}
	\caption{CAD drawing of the L'TES instrument. Infrared light enters the L'TES instrument through the Cassegrain telescope's 15.2 cm entrance aperture. The Aft Optics Plate holds the interferometer's moving mirror assembly, the beamsplitter, metrology laser and detector, internal calibration blackbody, and the infrared detector as well as various mirrors for directing the incoming light. The cube attached to the Aft Optics Plate is an alignment cube and is removed before launch. The four mounting flexures attach L'TES to the instrument pointing platform.   \label{fig:TES_CAD}}
\end{center}
\end{figure}

L’TES is a Fourier transform infrared point spectrometer (Figure \ref{fig:TES_CAD}) built at Arizona State University and Dr.~Phil Christensen is the Instrument Principal Investigator. L’TES has the same optical-mechanical design as OTES, including a 15.2~cm diameter Cassegrain telescope, a Michelson interferometer with chemical vapor deposited diamond beamsplitter, and an uncooled, deuterated L-alanine doped triglycine sulfate (DLATGS) pyroelectric detector.  L’TES has only small differences from the heritage instrument including removing a potential stray light path by modifying the telescope baffle and primary mirror inner diameter and improvements to the metrology laser system. An internal calibration cone blackbody target provides radiometric calibration. The L’TES instrument collects data from 6-75 $\mu$m and has a noise equivalent spectral radiance (NESR) of 2.3$10^{-8}$ W cm$^{-2}$ sr$^{-1}$/cm$^{-1}$ between 300 cm$^{-1}$ (7.4 \um) and 1350 cm$^{-1}$ (33 \um). For surfaces with temperatures greater than 75K, L’TES will determine the temperature with an accuracy of 2K. The 50$\%$ encircled energy of the instrument subtends 6.5 mrad. 

L’TES has one mode of taking data. It continuously collects interferograms (every 0.5, 1.0, or 2.0 sec) and transfers them to the spacecraft for storage before downlink. The instrument will start collecting data one day before closest approach, which is before the target fills the instrument’s field of view. The data collection will continue until one day after closest approach. The instrument regularly interleaves observations of an internal calibration mirror while taking science data. 

\begin{figure}[htbp]
	\centering
	\includegraphics[keepaspectratio,width=3 in]{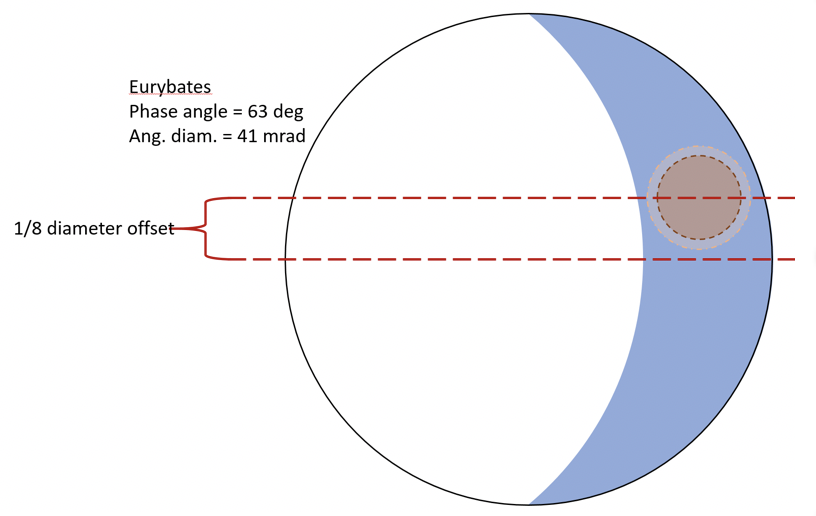}
	\caption{L'TES scan of a Trojan asteroid. To observe different local times of day, \Lucy\ will use the IPP to scan the L'TES instrument across the Trojan asteroid. The scan will start off the dark limb of the Trojan asteroid and progress across the lit hemisphere. For each of the \Lucy\ Trojan targets, there is a time when the L'TES field of view is smaller than the unilluminated region allowing a measure of the night side of the Trojan asteroid. There is a pointing uncertainty of 1/8 of the Trojan's diameter and that is \color{black} taken into account \color{black} when planning the timing of the scan.}
	\label{fig:ThermalScan}
\end{figure}

The L’TES instrument will measure the radiance of each Trojan asteroid at four locations at different local times of day with the additional requirement that one observation measures a location within 30$^{\circ}$ of the sub-solar point and another measures the un-illuminated surface (Figure \ref{fig:ThermalScan}). This observation is conducted by scanning the L’TES FOV across the asteroid. The measurements are converted into temperatures by fitting one or more blackbody curves (for known temperatures) to the measured radiance. These temperatures are inputs to models of thermal inertia, which are used to infer regolith properties.

\section{Spacecraft Subsystems Contributing Scientific Data} \label{sec:systems}

In addition to the scientific instruments, two spacecraft subsystems will be used to return scientifically valuable data. The Telecommunication subsystem will provide Doppler data to determine the mass of our targets. The Terminal Tracking Cameras (TTCam) which are part of the Guidance, Navigation and Control subsystem of the spacecraft will provide wide-field imaging that will contribute to the shape modeling of the Trojans. The concept of operations includes these systems in the encounter design. 

\subsection{TTCam}

\Lucy\ has two Terminal Tracking Cameras (TTCam) on board to provide autonomous pointing of the instrument pointing platform to each Trojan near closest approach. This system is part of the spacecraft's guidance and control system and will also collect monochromatic (475 - 625 nm bandpass) images that address our science objective to determine the shape of the Trojan asteroids. The wide field of view of the TTCams (11$^{\circ}$x8.2$^{\circ}$) allows imaging of the entire sunlit fraction of each of the Trojan asteroids during the entire flyby, unlike the high-resolution imager L'LORRI. One of the \Lucy\ shape objectives to derive stereo imaging requires observations separated by 15$^{\circ}$ to 25$^{\circ}$ of phase angle. While L'Ralph MVIC has higher spatial resolution, the use of TTCam images is preferred instead of L'Ralph because L'Ralph MVIC TDI imaging results in each line of the scan being a separate image to reconstruct the shape. Both L'Ralph MVIC and TTCam will be used to provide redundancy in the observing sequence.

\begin{figure}[ht!]
	\centering
	\includegraphics[keepaspectratio,width=4 in]{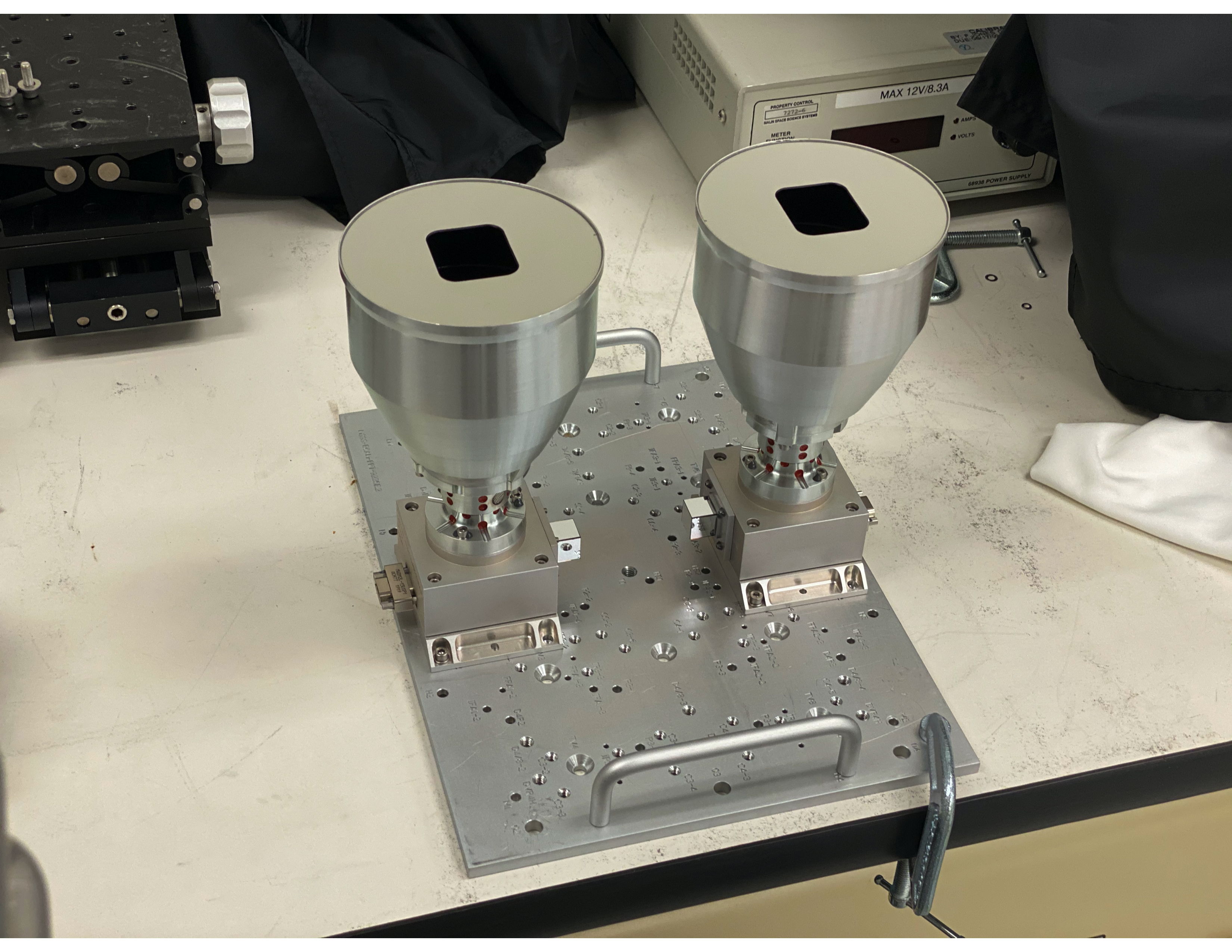}
	\caption{The Terminal Tracking Cameras in the lab at Malin Space Science Systems. 
	\label{fig:TTCams}}
\end{figure}

TTCam was built by Malin Space Science Systems and Dr.~James Bell is the Instrument Scientist. The instrument design is derived from the OSIRIS-REx Touch-And-Go Camera Suite \citep{Bos+2020}. The sensor is a \color{black} monochrome rolling-shutter \color{black} CMOS device with 2592x1944 active 2.2 $\mu$m pixels, or 2752 x 2004 pixels including masked dark pixels. The optical system has a focal length of 29.7 mm and a entrance aperture of 10 mm. Each pixel subtends 74.1 $\mu$rad. There are two terminal tracking cameras for redundancy. TTCam acquires 12-bit images that are companded onboard to 8-bits  using a square-root-like lookup table to preserve signal to noise and image dynamic range (e.g., \citet{Malin+2017}). Images are then losslessly compressed onboard for downlink using a standard first-difference Huffman encoding scheme. The primary imaging modes of TTCam are acquisition of in-flight calibration images with active and dark pixels, acquisition of high cadence tracking images for IPP pointing, and acquisition of moderate cadence images for science (shape modeling) within a few minutes around closest approach. The flight instruments are shown in Figure \ref{fig:TTCams}.

\subsection{Radio Science}

The Radio Science Investigation uses the spacecraft's X-band telecommunication system to measure the Doppler shift of the spacecraft. The mass of the Trojans imparts a delta-v on the spacecraft. The concept of operations to support the Doppler shift measurements entails measuring the Doppler signature of the spacecraft before and after the flyby to determine the change in velocity. There is a 7-hour baseline on either side of closest approach to collect Doppler data \color{black}(from E-9 hr to E-2 hr on approach and from E+2 hr to E+9 hr on departure)\color{black}. For 4 hours centered on closest approach the spacecraft's high gain antenna will not be pointed at Earth as a result of tracking the Trojan asteroid with the remote sensing instruments through the encounter. The only source of velocity disturbance on the spacecraft from 9 hours before encounter to 9 hours after encounter will only be the gravitational pull of the asteroid. The \Lucy\ mission will do any momentum management of the spacecraft's reaction wheels outside of this 18 hour window. 

The mass determination uncertainty is sensitive to the closest approach distance. Another consideration in the mass determination is the noise contribution of the solar wind. If the radio signal passes near to the Sun, then the Doppler measurement will be affected by the solar wind. Thus, the Sun-Earth-Probe (SEP) angle causes a degradation of the precision of the mass determination when the SEP angle is less than 10$^{\circ}$. For all our Trojan flybys, the SEP angle is greater than 20$^{\circ}$.

\section{Encounter Concept of Operations} \label{sec:conops}

A key consideration in designing the Trojan encounter concept of operations is to capture the required scientific data, in the critical time period near closest approach. The timing of each Trojan encounter establishes the heliocentric distance of the Trojan asteroid during the flyby as well as the solar phase angle (Sun-Trojan asteroid-spacecraft angle) on the approach and departure asymptotes. The apparent brightness of each Trojan asteroid during the flyby is a function  of the heliocentric distance, the Trojan asteroid's surface reflectivity which varies depending on the object's albedo and the solar phase angle. The apparent brightness of the Trojan targets throughout the encounter is important for planning the science observations and also the optical navigation imaging. For the first two encounters (Eurybates and Polymele), the spacecraft is approaching within 10$^{\circ}$ of the terminator. For the next two encounters (Leucus and Orus), the spacecraft is approaching from the unilluminated hemisphere. \textcolor{black}{The Trojan asteroids are not visible from Earth at such high phase angles so their phase curves are unknown}. Therefore, the \Lucy\ mission will take observations of the targets well in advance of the encounters (more than 1 year) to determine the apparent magnitude at phase angles that are greater than values achievable from Earth ($>$ 11$^{\circ}$). Given rotational light curves and phase curves at moderate phase angles, the mission will be able to better predict the apparent magnitude on approach.

\begin{deluxetable}{lccccc}[ht!]
	\tabletypesize{\scriptsize}
	\label{tab:EncounterCircumstances}
	\tablecolumns{6}
	\tablewidth{0pt}
	\tablecaption{Encounter Circumstances}
	\tablehead{	\colhead{Encounter} & \colhead{Date}& \colhead{Approach}& \colhead{Departure} & \colhead{Closest Approach} & \colhead{Flyby}\\ \colhead{} & \colhead{} & \colhead{Phase Ang. $^{\circ}$}& \colhead{Phase Ang. $^{\circ}$} & \colhead{Distance, km} &  \colhead{Velocity, km/s}}
	\startdata
	Eurybates & Aug. 12, 2027 & 81 & 99 & 1000 & 5.7  \\
	Polymele & Sept. 15, 2027  &  82 & 98 & 434 & 6.0 \\
	Leucus & April 18, 2028 & 104 & 76 & 1000 & 5.9  \\
	Orus & Nov. 11, 2028 & 126 & 54 & 1000 & 7.1 \\
	Menoetius & Mar. 2, 2033 & 56 & 124 & 1075 & 8.8 \\
	Patroclus & Mar. 2, 2033 & 56 & 124 & 1320 & 8.8 \\
	\enddata

\end{deluxetable}

\color{black}
\subsection{Closest Approach Distance Trades}
\color{black}

The closest approach distance for each target is determined by balancing the resolution needed for the highest resolutions images, the ability to detect the gravitational pull of the asteroid with Doppler tracking, and the time available during the flyby to accomplish observations at low-to-intermediate phase angles where surface shadowing is optimal for characterization of surface geology. The first two factors (high-resolution imaging and Doppler-shift measurements) benefit from smaller closest approach distances. At a closer distance the resolution of the images will be higher and the spacecraft will feel an increased gravitational force. However, as the closest approach distance is decreased, there is less time at lower phase angles needed for surface science, observations of craters and other geologic features at a phase angle of 30$^{\circ}$ to 60$^{\circ}$ which is desired to have shadows for geologic feature identification. 

Table \ref{tab:EncounterCircumstances} lists nominal closest approach distances. For most of our flybys, a closest approach distance of 1000 km from the center of the Trojan asteroid on the sunward side of the target satisfies our requirements. \color{black}The highest-resolution imaging requires identifying craters as small as 70 meters across \citep{Levison+2021}. We assume five pixels across a crater for identification and this results in a required resolution of 14 meters. A closest approach distance of less than about 1800 km allows for this resolution given L'LORRI images with a 15 $\mu$rad FWHM and a factor of two resolution improvement from deconvolution of multiple images. \color{black} 

For the smallest of the \Lucy\ targets, Polymele, the closest approach distance needs to be smaller (434 km) to determine its mass with an uncertainty of less than 25\%. \color{black} The science goal is to determine the density of the Trojans to an uncertainty of 33\% \citep{Levison+2021} and the volume and mass uncertainties were allocated to 25\% and 22\% respectively to fulfill this goal. \color{black}  

At the other extreme, the Patroclus-Meneotius binary (PMB) flyby can accommodate a larger flyby distance because the objects are large and the system mass is well known from their mutual orbit. For the PMB, the spacecraft is targeting an aim point of 1075 km relative to Meneotius. The resulting closest approach distance relative to Patroclus is 1320 km. The slightly larger closest approach, compared to the nominal value of 1000 km, allows more time for the spacecraft to observe both targets at intermediate phase angles (30$^{\circ}$-60$^{\circ}$).  

\color{black}
\subsection{Encounter Geometry}
\color{black}

\begin{figure}[ht!]
	\plotone{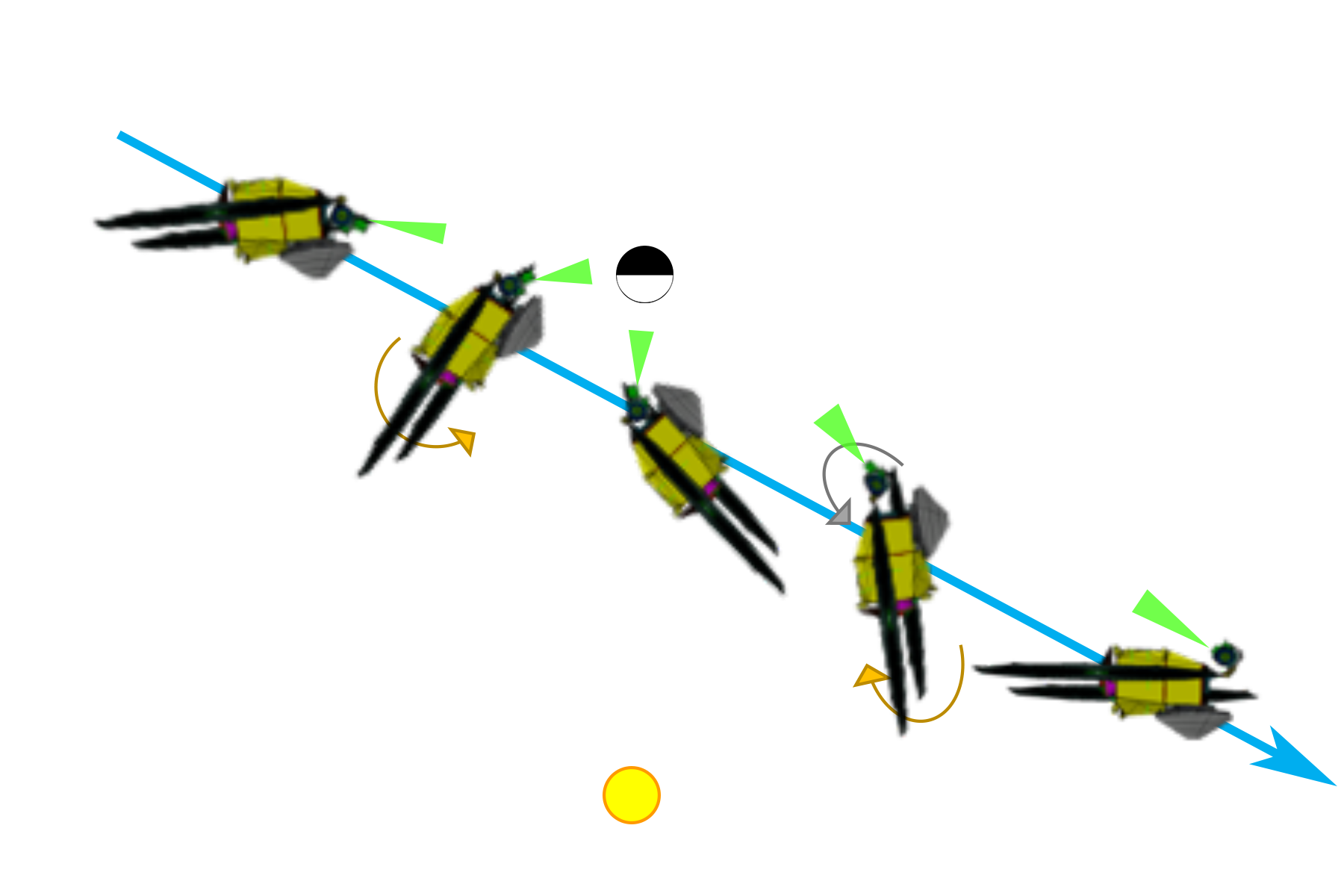}
	\caption{Spacecraft orientation during flyby when \Lucy\ approaches from phase angles greater than 90$^{\circ}$. The attitude of the \Lucy\ spacecraft relative to the Trojan asteroid is shown at 5 different time steps near closest approach. The Trojan asteroid is indicated by the black and white circle where the white region indicates the illuminated hemisphere. The green triangle indicates the viewing direction of the instruments on the IPP. The direction to the Sun is shown by the yellow circle. The \Lucy\ spacecraft rotates at less than 1$^{\circ}/s$ when tracking the Trojan asteroid. Near closest approach, the active area of the solar arrays are no longer being fully illuminated by the Sun. After closest approach (the 4th time step from the left) the spacecraft rotates back to have the active area of the solar arrays fully illuminated again. During this maneuver, the IPP rotates to keep the Trojan asteroid in the field of view of the instruments. Diagram is not to scale.   \label{fig:HeadFirst}}
\end{figure}

\begin{figure}[ht!]
	\plotone{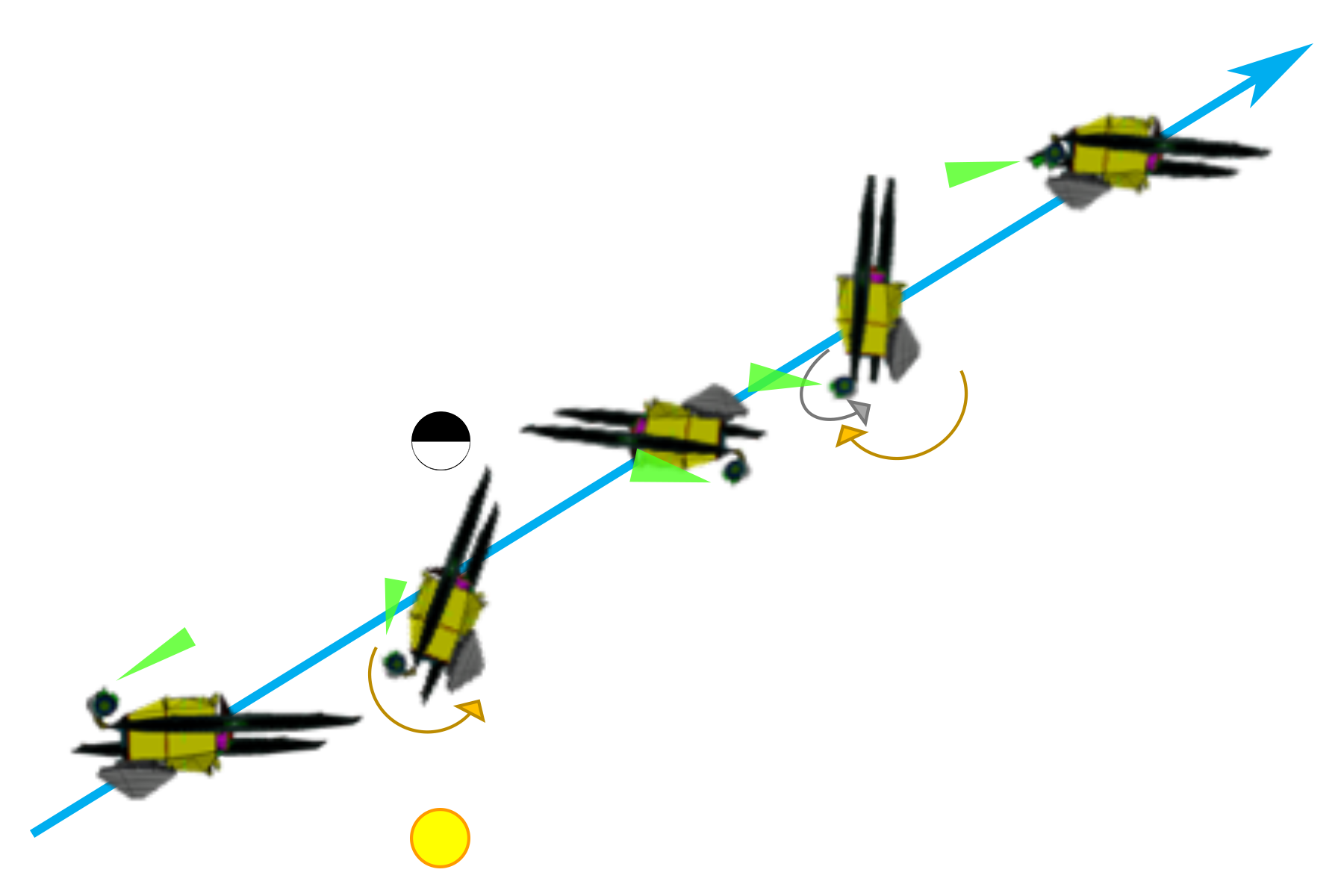}
	\caption{Spacecraft orientation during flyby when \Lucy\ approaches from phase angles less than 90$^{\circ}$. As in the previous figure, the attitude of the \Lucy\ spacecraft relative to the Trojan asteroid is shown at 5 different time steps near closest approach. In this case, the \Lucy\ spacecraft approaches the Trojan asteroid with the main engine first instead of the IPP. The general principal of using the spacecraft's rotation to keep the Trojan in view is the same in both scenarios. And in both scenarios, after closest approach the spacecraft rotates back to have the active area of the solar arrays fully illuminated again while the IPP rotates to keep the Trojan asteroid in the field of view of the instruments. \label{fig:FeetFirst}}
\end{figure}

\Lucy\'s scientific instruments and the terminal tracking cameras are mounted on an Instrument Pointing Platform (IPP) (Figure \ref{fig:IPP}) which is connected to the spacecraft by a two-axis gimbal. As shown in Figures \ref{fig:HeadFirst} and \ref{fig:FeetFirst}, the spacecraft tracks the Trojan asteroids by the combination of pitching the spacecraft and motion of the IPP. This allows the instruments to point to the Trojan asteroids throughout the flyby as the geometry changes quickly. 

The IPP has two planes of motion with different range limits. The outer gimbal is able to rotate by 199$^{\circ}$ and moves the IPP in encounter plane (the plane of the velocity vector and the Trojan asteroid) during the flyby. The inner gimbal has a range of motion of 24$^{\circ}$ and can move the IPP above and below the encounter plane. The two-axis gimbal provides the ability to point instruments at both members of the PMB system when the two objects are not co-aligned with the spacecraft's velocity vector. 

\color{black}The IPP enables nadir-pointed observations of the Trojan asteroids as well as the ability to construct a mosaic of images or scans across the Trojan asteroid. \color{black} 

\color{black}
\subsection{Optical Navigation and Science Data Collection}
\color{black}

Optical navigation begins at 60 days before closest approach (E-60 d). Science imaging begins at E-12 d \color{black} when we begin searching for satellites around our targets\color{black}. Most of the science data collection for the mission occurs within $\pm$4 days of closest approach. This 8-day time period is called the close-approach subphase. \color{black} Most of the data collection occurs within 4 days of closest approach because that is when the distance of the spacecraft is small enough to achieve the resolution requirements needed for the science (Table \ref{tab:Mapping}). Also three of the science requirements are to provide rotational coverage in panchromatic imaging, color imaging and infrared imaging spectra. The rotational period for the Lucy Trojan asteroids is about 100 hours or less \citep{Levison+2021} for all targets except Leucus which has a rotation period of 445 days \citep{Mottola+2020}. Therefore, the rotational coverage for all targets except Leucus can be achieved in the close-approach subphase. For Leucus, the rotational coverage will be conducted near closest approach and continuing onto the departure asymptote because that asymptote has a smaller phase angle than the approach asymptote (Table \ref{tab:EncounterCircumstances}. \color{black}.

During the close-approach subphase, the spacecraft fault protection system has enabled Auto Recovery mode. In this mode, if the spacecraft or instruments enter safe mode, the fault protection system will take action to clear the fault and rejoin the command sequence. The Auto Recovery mode is needed because significant closest approach science could be lost in the time it would take to identify a fault and fix the fault.

Most observations and actions on the spacecraft are commanded to execute at a given time. However, during the close-approach subphase most science observations will be initiated based on the range of the spacecraft to the Trojan asteroid target. At the beginning of this time period, the range is estimated on the basis of an on-board ephemeris. As the spacecraft approaches the target and the image of the target is resolved in the Terminal Tracking Cameras, the on-board terminal-tracking state estimation begins to provide an estimate of the Trojan's location relative to the spacecraft. This terminal tracking allows the \Lucy\ spacecraft to have updated knowledge of the target which allows for a more efficient observing \color{black} timeline. Instead of having to cover a large area with a large mosaic of images because the location of the Trojan relative to the spacecraft is unknown, the on-board terminal tracking system provides accurate knowledge of the location of the Trojan asteroid in the plane of the sky. \color{black} 

The Eurybates and Polymele encounters are separated by only 34 days. Before the Eurybates encounter occurs, the optical navigation campaign for Polymele will need to begin. For this reason, these two encounters will be planned as one. We will take advantage of the gimbal's ability to point to both Eurybates and Polymele without significant rotation of the spacecraft while interleaving post-Eurybates close-approach science observations and optical navigation observations on approach to Polymele.

\color{black}
The Science Operations Center (SOC) conducts the science observation planning and builds the command sequences for the operation of the science instruments on Lucy (L'LORRI, L'Ralph and L'TES). The SOC is located at the Southwest Research Institute in Boulder Colorado. The Mission Operations Center (MOC) is located at Lockheed Martin in Littleton Colorado. The Goddard Spaceflight Center in Greenbelt Maryland provides the Mission Operations Manager and support through its Space Science Mission Operations (SSMO) organization. 
\color{black}

\section{Conclusions}

The \Lucy\ mission's concept of operations has been designed to carry out the science goals of the mission given the capabilities of spacecraft systems and the instrument payload and the constraints of the planned spacecraft trajectory. The concept of operations is testable using a hardware-in-the-loop simulator and with the \Lucy\ spacecraft following instrument integration. Science sequences have been developed that demonstrate that all Level 1 science objectives are achievable. 

The \Lucy\ science team, in cooperation with the planetary science community, will continue ongoing efforts to improve knowledge of the \Lucy\ mission targets through Earth-based observations of occultations, light curves, spectroscopy and deep-imaging satellite searches. New information will be incorporated into the science planning for each of our targets as appropriate. After launch, in-flight instrument performance will also be folded into the \Lucy\ encounter plans.


\acknowledgments

The authors thank the entire Lucy mission team for their hard work and dedication.

%








\bibliography{LucyMissionOverview}{}
\bibliographystyle{aasjournal}



\end{document}